\documentstyle{article}

 \def\vt{t\kern-0.22em\raise.18ex\hbox{\char'47}\lower.18ex\hbox{}\kern-0.08em}
 
\newtheorem{th}{Theorem}[section]

\newtheorem{ex}{Example}[section]

\newtheorem{rem}{Remark}[section]

\newcommand{\old}[1]{{}} 
\newcounter{obr}
\newcounter{tabul}
\begin{document}
\title{Bounds  for the Clique Cover Width  of Factors of the Apex Graph 
of the Planar Grid  
}
\author{Farhad Shahrokhi\\
Department of Computer Science and Engineering,  UNT\\
farhad@cs.unt.edu
}

\date{}
\maketitle
\thispagestyle{empty}
\date{} \maketitle

%\baselineskip=17.6pt

%%%%%%%%%%%%%%%%%%%%%%%%%%%%%%%%%%%%%%%%%%%%%%%%%%%%%%%%%%%%%%%%%%%%%%%%%%%%%%%% 
\begin{abstract}

The  {\it clique cover width} of  $G$,
denoted by $ccw(G)$, is the minimum  value of the  bandwidth of
all  graphs that are obtained by contracting  the cliques in a clique cover of $G$ into a single vertex.    
For $i=1,2,...,d,$ let $G_i$ be
a graph with $V(G_i)=V$, and  let $G$ be a graph with $V(G)=V$ and 
$E(G)=\cap_{i=1}^d(G_i)$, 
then we  write   $G=\cap_{i=1}^dG_i$ and call each $G_i,i=1,2,...,d$ a factor of $G$. 
The case where $G_1$ is chordal, and  
for $i=2,3...,d$  each factor $G_i$ has a ``small"  $ccw(G_i)$,   is well 
studied due to applications. 
Here we show a negative result. Specifically, let ${\hat G}(k,n)$ be  
the graph obtained by joining a set of $k$ apex  vertices of degree $n^2$ to all vertices of an $n\times n$ grid, and then adding  some possible edges 
among these $k$ vertices.
We prove that if  ${\hat G}(k,n)=\cap_{i=1}^dG_i$, with $G_1$ 
being chordal, then, 
$max_{2\le i\le d}\{ccw(G_i)\}=
\Omega({n^{1\over d-1}})$. 
Furthermore, for $d=2$, we  
construct a chordal graph $G_1$ 
and a graph $G_2$ with $ccw(G_2)\le {n\over 2}+k$ so that ${\hat G}(k,n)=G_1\cap G_2$. 
Finally, let ${\hat G}$ be the clique sum graph of  ${\hat G}(k_i, n_i)$,  where for $i=1,2,...t$, the  underlying grid is $n_i\times n_i$ and the sum is taken at apex vertices. 
Then, we show ${\hat G}=G_1\cap G_2$, where, $G_1$ is chordal and $ccw(G_2)\le \sum_{i=1}^t(n_i+k_i)$. 
The implications and applications of the results are discussed, including addressing a recent question of David Wood. 
%A universal representation theorem is derived that asserts 
%$G=H_1\cap H_2$, for any $G$, where $H_1$ is chordal, and $CCW(H_2)$ is bounded by  
%parameters that depend  on the properties of the tree decompositions of $G$.     

\end{abstract}

\section{Introduction}
In this paper, $G=(V(G), E(G))$ denotes  an undirected  graph.
 A {\it chordal graph} is a graph with no chordless 
cycles \cite{Go}. 
An {\it interval graph} is the intersection graph of the intervals on the real line
\cite{Tr}. A {\it comparability  graph} is a graph whose edges have a transitive orientation.  An {\it incomparability graph} \cite{G1} is the complement of a comparability graph.   
A clique cover $C$ in $G$ is a partition of $V(G)$ into cliques.  
Throughout this paper, we will  write
$C=\{C_0,C_1,...,C_t\}$ to indicate that $C$ is an ordered set of cliques in $G$.
Let $L=\{v_0,v_2,...,v_{n-1}\}$ be a linear ordering of vertices
in $G$. The {\it width} of $L$, denoted by $w(L)$, is
$max_{v_iv_j\in E(G)}|j-i|$.
The  {\it bandwidth} of $G$ 
denoted by $bw(G)$, is the smallest  width of all linear orderings
of $V(G)$ \cite{CC}. 
For a clique cover $C=\{C_0,C_1,...,C_t\}$, in $G$, let
the {\it width}  of $C$, denoted by $w(C)$, denote
$\max\{|j-i||xy\in E(G), x\in C_i,y\in C_j,C_i,C_j\in C\}$.
The {\it clique cover width } of $G$ denoted by $ccw(G)$, is the smallest width
all ordered clique covers in $G$\cite {Sh1}. 
Note that $ccw(G)\le bw(G)$. 
It is easy to verify that any 
graph with 
$ccw(G)=1$ is an incomparability graph.

Let $d\ge 1$, be an integer, and for $i=1,2,...,d$ let $G_i$ be 
a graph with $V(G_i)=V$, and  let $G$ be a graph with $V(G)=V$ and $E(G)=\cap_{i=1}^d(G_i)$.
Then   we say $G$ is the {\it edge intersection graph} of $G_1,G_2,...,G_d$, and write $G=\cap_{i=1}^dG_i$. In this setting,  each $G_i, i=1,2,...,d$ 
is a {\it factor} of $G$.
Roberts \cite{Ro} introduced the {\it boxicity} and {\it cubicity} of a 
graph $G$ as the smallest integer $d$ so that $G$ is the edge 
intersection graph
of $d$ interval, or unit interval graphs, respectively. For more recent work see
\cite{Sul, Sul2, Sul3}.   

For integers $d\ge 2$ and $w\ge 1$, let ${\cal C}(d,w)$ be the 
class  all graphs 
$G$ so that $G=\cap_{i=1}^d G_i$,  where $G_1$ is chordal, and for 
$i=2,3,...,d, ccw(G_i)\le w$.  
A function that assigns non-negative values
to subgraphs of $G$, is called a  measure, if the following hold.
$(i)$,  $\mu(H_1)\le \mu(H_2)$, if $H_1\subseteq H_2\subseteq G$,
$(ii)$ $\mu(H_1\cup H_2)\le \mu(H_1)+\mu(H_2)$,
if  $H_1,H_2\subseteq G$,
$(iii)$  $\mu(H_1\cup H_2)=\mu(H_1)+\mu(H_2)$, if there are no edges between $H_1$ and $H_2$.

Motivated by the earlier results of Chan \cite{Chan} and Alber and Fiala 
\cite{AF} on geometric separation of 
fat objects (with respect to a measure),  we have  derived in \cite{Sh} 
a general combinatorial separation Theorem 
whose statement without a proof
was announced in \cite{Sh1}.

\begin{th}\label{t1}(\cite{Sh1, Sh})

{\sl Let $\mu$ be a measure on  $G=(V(G),E(G))$, and  let $G_1,G_2,...,G_p$
 be  graphs  with $V(G_1)=V(G_2)=,...,V(G_p)=V(G)$, $d\ge 2$ and
$E(G)=\cap_{i=1}^p E(G_i)$ so that $G_1$ is chordal. Then there is
a vertex separator $S$ in $G$ whose removal separates $G$
into two subgraph so that each subgraph has a measure of at most
 $2\mu(G)/3$. In addition,
the induced graph of $G$ on $S$   can be covered with
at most $O(2^d{l^*}^{d-1\over d}{\mu(G)}^{d-1\over d})$ many  cliques
from $G$, where $l^*=max_{2\le i\le d}ccw(G_i)$.
}

\end{th}

A number of  geometric applications of Theorem \ref{t1}  
were announced in \cite{Sh1}, however, in all these cases $G_1$ 
happened to be an interval graph, and hence the full power of the separation 
theorem was not utilized. We remark that a slightly stronger version of the 
above Theorem has been  proved in \cite{Sh}, where  $l^*={\Pi_{i=2}^dccw(G_i)}^{1\over d-1}$.

Recall that a   {\it tree decomposition} \cite{RS,bod}
of a graph $G$ is a pair $(X,T)$ 
where $T$ is  a tree, and $X=\{X_i|i\in V(T)\}$ is a family of
subsets of $V(G)$, each called a bag, so that the following hold:

$\bullet$ $\cup_{i\in V(T)}X_i=V(G)$

$\bullet$ for any $uv\in E(G)$, there is an $i\in V(T)$ so that
$v\in X_i$ and $u\in X_i$.

$\bullet$ for any $i,j,k\in V(T)$, if $j$ is on the path from
$i$ to $k$ in $T$, then $X_i\cap X_k\subseteq X_j$.

Width of $(X,T)$, is the size of largest  bag minus 1. Treewidth  of $G$, or 
$tw(G)$ is the smallest width, overall tree decompositions of $G$.   In an attempt to  explore the  structure of class ${\cal C}(2,w)$ we proved  
Theorem \ref{t2} in \cite{Sh3}.

\begin{th}\label{t2}(Universal Representation Theorem:\cite{Sh3})
{\sl
Let $G$ be a graph and let $L=\{L_1,L_2,...,L_k\}$ be  a
partition of vertices, so that for any  $xy\in E(G)$, either  $x,y\in L_i$
where $1\le i\le k$, or,
$x\in L_i, y\in L_{i+1}$, where,   $1\le i\le k-1
$.  Let $(X,T)$ be a tree decomposition of $G$. Let
$t^*=\max_{i=1,2,...,k}\{|L_i\cap X_j||j\in V(T)\}$. (Thus, $t^*$
is the largest number of vertices in  any element of $L$  that appears in any 
bag of $T$).
Then, there is a chordal graph $H_1$ and a graph $H_2$  with $ccw(H_2)\le 2t^*-1$
so that  $G=G_1\cap G_2$.
}
\end{th}

Noting that  for any planar graph $G$, the parameter $t^*$ is at most 4, 
it follows that  any planar $G$ is in class ${\cal C}(2,7)$ \cite{Sh3} (with $G_1$ being a chordal graph).   
Consequently, 
the planar separation theorem \cite{LT}follows from 
Theorem \ref{t1}. 
We further speculated in \cite{Sh3}   that similar  results would hold for graphs drawn of surfaces, and graphs excluding specific minors.

Concurrent with our work in \cite{Sh3}, in 2013, and independently from us, Dujmovi\'c, Morin,   and Wood \cite{DMW}, formalized the notation of $t^*$, and introduced  a  parameter  called the {\it layered tree width}, or $ltw(G)$,  
which is the minimum value of $t^*$, over all tree decompositions and layer partitions of $G$.  
They provided  significant applications 
 in graph theory and graph drawing, under the requirement that 
$ltw(G)$ is bounded by a constant.
Among other results, it was shown in \cite{DMW}  that
$ltw(G)\le 2g+3$, for any genus $G$ graph, and as a byproduct   some well known 
separation theorem followed  with improved multiplicative  constants. 
Dujmovi\'c, Morin, and Wood further classified  those graphs $G$ with bounded   
$ltw(G)$ to be the class of  $H$ minor free graphs, for a fixed  apex graph  $H$.   
($H$ is an apex graph, if  $H-\{x\}$ is a planar graph for some vertex $x$.) 
Combining this result and   the  Universal Representation Theorem, it can be
concluded that for any fixed apex graph $H$, there is an integer $w(H)$, 
so that any $H-$minor free graph $G$, is in 
class ${\cal C}(2,w(H))$.
Using the terminology in the abstract, let   ${\hat G}(1,n)$, be the 
graph obtained
from  $n\times n$ grid  by adding a new vertex  of  degree $n^2$ which is adjacent to all vertices of the grid.  Then, although this graph  not have $K_6$ as a minor,  it  does not satisfy the  requirement for the usage of the  
framework in \cite{DMW},  since $K_5$ is not 
planar.  
Specifically,  as noted in \cite{DMW},  $ltw({\hat G}(1,n))=\Omega(n)$, 
and hence the layered treewidth method is not applicable. Note that the 
Universal Representation Theorem also fails to  show that ${\hat G}(1,n)$ is 
in ${\cal C}(2,w)$ for any constant $w$. David Wood \cite{Wood}
in private communications  raised the following question:
Might it  be  true that  for every $H$-minor-free graph $G$, 
${G}\in {\cal C}(2,w(H))$, for some constant $w(H)$ 
depending on  $H$?  Particularly, he inquired if this is true for ${\hat G}(1,n)$.
Now, let ${\hat G}(k,n)$  be 
the graph obtained  by joining  a set $X$ of $k$ (apex) vertices  of degree $n^2$ to all vertices of an $n\times n$ grid, with possible addition of edges among these vertices in $X$. 
In Section  two we prove that if ${\hat G}(k,n)=\cap_{i=1}^dG_i,$ where $G_1$ is chordal, 
then 
$max_{2\le i\le d}\{ccw(G_i)\}=
\Omega(n^{1\over d-1})$,  
 and hence answer Wood's  question in negative, by letting $k=1$ and $d=2$.  
In the positive direction, for  $d=2$, we show  ${\hat G}(k,n)=G_1\cap G_2$, so that 
$G_1$ is chordal and $ccw(G_2)\le {n\over 2}+k$, where the upper bound of ${n\over 2}+k$ is small enough for the effective application of  
the separation Theorem \ref{t1}.    
We extend this result to clique sum graphs. Specifically,  let ${\hat G}$ be the clique sum graph of  ${\hat G}(k_i,n_i), i=1,2,...t$, where the  underlying grid is $n_i\times n_i$ and the sum is taken at apex vertices. 
Then, we show ${\hat G}=G_1\cap G_2$, where, $G_1$ is chordal and $ccw(G_2)\le \sum_{i=1}^t(n_i+k_i)$.

\section{Main Results}

\begin{th}\label{t3}
{\sl 
Let ${\hat G}(k,n)$  be 
the graph obtained  by joining  a set $X$ of $k$ vertices  of degree $n^2$ to all vertices of an $n\times n$ grid, with possible addition of edges among these vertices. 
Then, the following hold. 

\noindent $(i~)$ If ${\hat G}(k,n)=\cap_{i=1}^dG_i$, where $G_1$ is chordal,   
then,  
$\max_{2\le i\le d}\{ccw(G_i)\}=
\Omega(n^{1\over d-1})$.  

\noindent $(ii~)$ There is a chordal graph $G_1$ and a graph $G_2$ with $ccw(G_2)\le {n\over 2}+k$, so that ${\hat G}(k,n)=G_1\cap G_2$.}
\end{th}

{\bf Proof.} For $(i)$,  let $H$ be the $n\times n$ planar  grid. Robertson and Seymour 
have shown that any chordalization of $H$ has a clique size $n/c$, for a constant $c$. 
Since $G_1$ restricted to $H$ is chordal,  it follows that, $G_1$ induced to 
$H$, must have  a clique $S$ of size $n/2c$. It follows that 
the subgraph of ${\hat G}(k,n)$ induced on $S$ has a independent set $S'$ of 
size at least $n/2c$. Now let $G'=\cap_{i=2}^{d}G_i$, and observe that 
$S'$ must also be an independent set in $G'$, since ${\hat G}(k)=G_1\cap G'$, 
and $S'$ is a clique in $G_1$.  Let ${\hat S}=S\cup \{x\}$ for some $x\in X.$
 Next for $i=2,3,...,d$, assume that 
$C_i$ is a clique cover in $G_i$ with $w(C_i)=ccw(G_i)$ and  let $B_i$ be the 
restriction of $C_i$ to $\hat S$, and let $2\le j\le d$ so that  
$|B_j|=max_{2\le i\le d}\{|B_i|\}$.
Note that   $|(B_j)|\le 2w(B_j)$, since $x$ is adjacent to all
vertices in $S'$. Now if $|(B_j)|\ge 
{n^{1\over d-1}\over {(2c)}^{1\over {d-1}}}$, 
then the claim follows.
So assume that 
$|(B_j)|
<{n^{1\over d-1}\over {(2c)}^{1\over {d-1}}}$, then  
$S'$ can be covered with strictly less than 
${|B_j|}^{d-1}={n\over 2c}$ cliques in  $G'$ which is a contradiction,
since $S'$ is independent in $G'$ and   $|S'|\ge {n\over 2c}$.   

For $(ii)$, take the vertices of the grid $H$ in every two 
consecutive rows,  and make them into one single clique, by the  addition   
of edges. This way we get a graph $H'$ which is a unit interval graph. Now make the set $X$ a clique, and   add this clique and all $k.n^2$ edges 
between $X$ and  vertices of $H$ to edges of $H'$ to  obtain   a graph $G_1$ which can be shown to be  chordal.  
To construct $G_2$, take any column of $H$, and make all vertices in this 
column into 
one single clique. This way, we obtain a graph $I$  with  a clique cover,  
$O=\{C_1,C_2,...,C_n\}$, where the vertices in each clique in $O$ are the vertices in a column of $H$. It is easily seen that $ccw(I)=1$. To obtain
$G_2$, add to $I$ all vertices in  $X$ and edges incident to them. 
To complete the proof 
place each vertex $x\in X$ as a clique between   $C_{n\over 2}$, and 
$C_{{n\over 2}+1}$.  Observe that ${\hat G}(k,n)=G_1\cap G_2,$ and 
$ccw(G_2)\le {n\over 2}+k$.  $\Box$ 

\begin{rem}\label{CON}
{\sl Let $G, |V(G)=N|$ be an  $H$ minor free graph, where $H$ is a fixed graph.
It is known that $tw(G)=O(\sqrt{N})$ \cite{Gro},  consequently by the Universal Representation 
Theorem,  $G=G_1\cap G_2$, where $G_1$ is chordal and $ccw(G_2)=O({\sqrt{N}})$. 
Moreover, the upper bound of $O({\sqrt{N}})$  is tight, 
since ${\hat G}(1,n)$ has $N=n^2+1$ vertices, is $k_6$ minor free and,  
by Part $(i)$ in Theorem \ref{t3}, if ${\hat G}(1,n)=G'_1\cap G'_2$, where 
$G'_1$ is chordal, then $ccw(G'_2)=\Omega(n)$.    
 
}
\end{rem}

\begin{rem}\label{r1}
{\sl By Part $(i)$, ${\hat G}(k,n)\notin {\cal C}(d,w)$, for any constants $w$ and 
$d$.  Nonetheless,  the upper bound for $ccw(G_2)$ in Part $(ii)$  of 
Theorem \ref{t3} is sufficient 
to use 
Theorem \ref{t1}  and show that 
   ${\hat G}(k,n), |V({\hat G}(k))|=N$ has  a vertex separator  
of size $O(N^{1-\epsilon})$, with the splitting ratio $1/3-2/3$,  where $N$, is the number of vertices in ${\hat G}$, as long as  
$k=O(n)$. The bound on the separator size is sufficient for many 
algorithmic purposes, although it may not be the best possible. 
}
\end{rem}

%\section{Clique Sum Graphs}

Let $G_1$ and $G_2$ be graphs so that $V(G_1)\cap V(G_2)$ is a clique in both
$G_1$ and $G_2$. Then, the {\it clique sum} of $G_1$ and $G_2$, denoted by $G_1\oplus G_2$ is a graph $G$ with $V(G) = V(G_1)\cup V(G_2)$, and $E(G) = E(G_1)\cup E(G_2)-E'$,
where $E'$ is a subset of edges (possibly empty) in the  clique induced by
$V(G_1)\cap V(G_2)$. The clique sum of more than two graphs is defined
iteratively,  using the definition for two graphs.
Let $G=G_1\oplus.... G_2\oplus ...\oplus G_k$ then we write $G=\oplus_{i=1}^kG_i$.
Clique sums are intimately related to the concept of the tree width and tree
decomposition; Specifically, it is known that if the tree widths of $G_1$ and
$G_2$
are at most $k$, then, so is the tree width of $G_1\oplus G_2$.

\begin{th}\label{t5}
{\sl 
For $i=1,2,...,t$, let ${\hat G}(k_i,n_i)$ denote the apex graph 
of an $n_i\times n_i$ grid, with an apex  set $X_i, |X_i|=k_i$, so that 
every vertex in $X_i$ is adjacent to all vertices of the $n_i\times n_i$ grid. 
Let 
${\hat G}=\oplus_{i=1}^t{\hat  G}_{(k_i,n_i)}$, where the clique sum is only taken at apex sets.
Then,  ${\hat G}=G_1\cap G_2$, where, $G_1$ is chordal and $ccw(G_2)\le \sum_{i=1}^t(n_i+k_i)$. 

}
\end{th}
{\bf Proof.} We use the construction in part $(ii)$ of Theorem \ref{t3}.   
Thus, for $i=1,2,...,t$, there is a chordal graph $G^i_1$ and a graph $G^i_2$ 
with $ccw(G^i_2)\le {n_i\over 2}+k_i$, so that ${\hat G}(k_i,n_i)=G^i_1\cap G^i_2$.
Let $G_1=\oplus_{i=1}^t G^i_1$, and $G_2=\oplus_{i=1}^tG^i_2$. Then, 
${\hat G}=G_1\cap G_2$. It is easy to verify that $G_1$ is chordal. 
To verify the claim concerning $ccw(G_2)$, we use 
the construction for general graphs, in \cite{Sh4}, in a simpler setting.   
Particularly, 
let $O_i$ be the clique cover 
for $G^i_2, i=1,2...,t$, where  each vertex $x_i\in X_i$ is 
represented  as a  clique and  has already  been placed in the ``middle" 
of $O_i$. When   
taking the sum of ${\hat G}_{(k_i,n_i)}$ and ${\hat G}_{(k_{i+1},n_{i+1})}, i=1,2,..,t-1$,  
we identify the vertices of the clique in 
$X_{i+1}$  with the clique in $X_i$, first, and then place 
the cliques in $O_{i+1}$ (in the same order that they appear in $O_{i+1}$) 
so that each clique in $O_i$ appears immediately to the left of a clique in $O_{i+1}$. 
$\Box$

We finish this  section  by exhibiting  graphs in ${\cal C}(2,1)$ with  
arbitrary large layered tree width. For these graphs, 
Theorem \ref{t1} is  applicable, directly,  
but the  Universal Representation Theorem would fail to be effective.

\begin{ex}\label{e3} 
{\sl 

\noindent  $~(i)$ For positive  integers $n,k$,  there is an incomparability graph 
$G\in {\cal C}(2,1)$ of diameter $\Theta(k)$   on $\Theta(nk)$ vertices   
with $ltw(G)=\Omega(n/k)$. 

\noindent $~(ii)$ For positive integers $n,k$, there is a graph $G\in {\cal C}(2,1)$ 
with $n^2k$ vertices  of diameter  $\Theta({n})$ so that 
$ltw(G)=\Omega(k/n)$.  
}
\end{ex}

For $(i)$, let $G$ be a graph with $ccw(G)=1$, or a unit incomparability graph on $N=n.k$ vertices, of diameter $k+2$, where each maximal clique has $n$ vertices. 
So there is a clique cover $\{C_1,C_2,...,C_k\}$ of $G$ so that any edge is either in the same clique or joins two consecutive cliques.  
Let $G_2=G$. Now let $G_1$ be a graph which is obtained  by adding all 
possible edges between two  consecutive cliques of $G$, then $G_1$ is an interval graph.   
Note that, $G=G_1\cap G_2$,  and $n\le tw(G)\le ltw(G)k$. 

For $(ii)~$, let $H$ be $n\times n$ grid embedded in the plane. 
To obtain $G$ replace any vertex $x\in V(G)$ by a clique $c_x$ of $k$ vertices. 
Now for any $x,y\in V(H)$ with $xy\in E(H)$ place $k^2$ edges joining every vertex in $c_x$ to very vertex in $c_y$  in $G$. 
To obtain $G_1$, take every two consecutive rows in $H$, and make all 
vertices of $G$ in them into one single 
clique in $G_1$. It is easy to  verify that $G_1$ is an interval graph. 
To construct $G_2$, take any column of $H$, and make all vertices of $G$ in them into one single clique in $G_2$. It can be verified that $ccw(G_2)=1$, and that
${G}=G_1\cap G_2$. $\Box$

\end{document}